\documentclass[format=acmsmall, review=false, screen=true]{acmart}

\usepackage{booktabs} 

\usepackage[ruled]{algorithm2e} 

\SetAlFnt{\small}
\SetAlCapFnt{\small}
\SetAlCapNameFnt{\small}
\SetAlCapHSkip{0pt}
\IncMargin{-\parindent}

\acmJournal{PACMHCI}
\acmVolume{2} 
\acmNumber{CSCW} 
\acmArticle{200} 
\acmMonth{11} 
\acmPrice{15.00}
\acmYear{2018}
\copyrightyear{2018}

\setcopyright{acmlicensed}

\acmDOI{10.1145/3274469}

\received{April 2018}
\received[revised]{July 2018}
\received[accepted]{September 2018}

\begin{document}

\title[User Perceptions of Smart Home IoT Privacy]{User Perceptions of Smart Home IoT Privacy}

\author{Serena Zheng}
\affiliation{%
  \institution{Princeton University}
  \streetaddress{35 Olden Street}
  \city{Princeton}
  \state{NJ}
  \postcode{08540}
  \country{USA}}
\email{serenaz@alumni.princeton.edu}

\author{Noah Apthorpe}
\affiliation{%
  \institution{Princeton University}
  \streetaddress{35 Olden Street}
  \city{Princeton}
  \state{NJ}
  \postcode{08540}
  \country{USA}}
\email{apthorpe@cs.princeton.edu}

\author{Marshini Chetty}
\affiliation{%
  \institution{Princeton University}
  \streetaddress{35 Olden Street}
  \city{Princeton}
  \state{NJ}
  \postcode{08540}
  \country{USA}}
\email{marshini@cs.princeton.edu}

\author{Nick Feamster}
\affiliation{%
  \institution{Princeton University}
  \streetaddress{35 Olden Street}
  \city{Princeton}
  \state{NJ}
  \postcode{08540}
  \country{USA}}
\email{feamster@cs.princeton.edu}

\begin{abstract}
Smart home Internet of Things (IoT) devices are rapidly increasing in popularity, with more households including Internet-connected devices that continuously monitor user activities. In this study, we conduct eleven semi-structured interviews with smart home owners, investigating their reasons for purchasing IoT devices, perceptions of smart home privacy risks, and actions taken to protect their privacy from those external to the home who create, manage, track, or regulate IoT devices and/or their data. We note several recurring themes. First, users' desires for convenience and connectedness dictate their privacy-related behaviors for dealing with external entities, such as device manufacturers, Internet Service Providers, governments, and advertisers. Second, user opinions about external entities collecting smart home data depend on perceived benefit from these entities. Third, users trust IoT device manufacturers to protect their privacy but do not verify that these protections are in place. Fourth, users are unaware of privacy risks from inference algorithms operating on data from non-audio/visual devices. These findings motivate several recommendations for device designers, researchers, and industry standards to better match device privacy features to the expectations and preferences of smart home owners.
\end{abstract}

%
%


%
%


\maketitle

\renewcommand{\shortauthors}{}

\section{Introduction}

Smart home Internet of Things (IoT) devices have a growing presence in consumer households.
Learning thermostats, energy tracking
switches, video doorbells, smart baby monitors, and app- and voice-controlled
lights, shades, and speakers are all increasingly available and affordable. 
These connected devices use embedded sensors and the Internet
to collect and communicate data with each other and their users, seamlessly
integrating the physical and digital worlds inside the home.

As smart home IoT devices become more popular and upload more private
data to the cloud, questions about user privacy regarding external entities who create, manage, track, or regulate these devices and/or their data arise:
What data do smart home IoT devices collect? Where is the data stored? Who has
ownership and access to the data? How is the data used? The answers
to these questions are often unclear, and the nascent
market for smart home devices has minimal regulations
or standards. As IoT devices become more ubiquitous, clarifying their privacy implications
is of utmost importance so that users are made aware of IoT related privacy risks and that these risks can be minimized without placing the burden of doing so solely on the users themselves. 

Researchers have studied IoT privacy challenges \cite{Arabo,Bugeja}
and proposed potential solutions \cite{Perera,Jacobsson, Nurse} to protect
user privacy. Such solutions are most effective when they account for the attitudes and awareness of end users. Our work builds on previous 
studies investigating user interactions with IoT devices~\cite{Worthy,Udoh,McReynolds,Zeng}, grounding the conversation about IoT privacy with empirical evidence about users' privacy perceptions with respect to entities external to the home, including device manufacturers, advertisers, Internet service providers (ISPs), and governments.
We focus on external entities because they determine the data collection capabilities of smart home devices, have access to data from large numbers of households, and are well-known privacy threats in other contexts, such as cloud services~\cite{Greenwald}, phone call records~\cite{Greenwald-phone}, and social media~\cite{Jung}.
By continuing to monitor
users' values and expectations in response to rapidly evolving smart home technology, we can help inform
better design practices and policies to ensure IoT devices are privacy-preserving without extensive user intervention.

We conducted eleven semi-structured interviews with smart home owners 
in the United States about their long-term experiences living with
IoT devices. The interviews were based on several focused questions and expanded into wide-ranging discussions 
about privacy awareness, privacy concerns, and device use cases. 
The participants owned a wide variety of IoT devices and shared a broad range of 
experiences of how these devices have impacted their lives. 
They also expressed a range of knowledge and personal opinions about
privacy concerns, including intentional purchasing and device interaction decisions made 
based on privacy considerations. 
Several common threads emerged across interviews.
We have distilled these recurring themes as our primary results and used them to generate
recommendations for IoT device designers, researchers, industry standards, and consumer incentives. 

We recognize that notions of privacy are situated in the everyday contexts of home routines, including the relationships between household members, and that our study captures only a snapshot of the privacy preferences and expectations of users based on their knowledge and expectations of entities not typically involved in day-to-day home life. This situated nature of privacy prevents us from unpacking all factors influencing IoT device design and privacy in one paper. Instead, our study provides valuable evidence of various perceptions of IoT devices and the entities who directly influence what data these devices collect, track, store, and transfer from users who have incorporated these devices into their daily lives.  Our study therefore makes a contribution to the CSCW literature on home IoT devices that can inform the design of  privacy-preserving and privacy-enhancing domestic technologies.  

\textbf{Convenience and connectedness are priorities for 
smart home owners, and these values dictate privacy opinions and behaviors directed at external entities that create, manage, track, or regulate IoT devices and their data} (Section~\ref{sec:findings-conv}).
Convenient features provided by IoT devices were 
the most frequently cited reason for adopting IoT technology and for 
disregarding concerns about personal privacy risks.
These results support previous work on user behaviors in contexts other than smart home IoT~\cite{Kang, Acquisti, Wang, Demiris,Townsend}, and indicate that convenience remains a primary justification for sacrificing privacy for IoT device owners. 

 \textbf{User opinions about who should have access to their smart home data
depend on notions of perceived benefit from entities external to the home} (Section~\ref{sec:findings-benefit}).
Participants viewed sharing data with a particular entity as more permissible if they believed 
that it would result in tangible benefits to themselves or their families. 
Device manufacturers providing software updates and new features were the least concerning 
recipients of smart home data.
Participants had mixed opinions about sharing data with advertisers and government entities,
depending on their opinions about the benefits of personalized advertising and the potential for 
local governments to improve services based on smart home data analysis.
 ISPs were uniformly distrusted with smart home data; 
participants  did not believe that ISPs could provide any benefits to smart home owners and
expressed strong privacy concerns about ISPs accessing their information. These results provide supporting qualitative evidence for previous survey findings about ISP privacy concerns~\cite{Apthorpe-CI} and provide evidence of user opinions of targeted advertising in a new context.

\textbf{User assumptions about privacy protections are contingent 
on their trust of IoT device manufacturers} (Section~\ref{sec:findings-trust}).
Brand familiarity and reputation were driving factors in IoT device purchasing decisions. 
Participants generally believed that devices made by well-known companies, both traditional 
technology companies and home appliance companies, were more likely to successfully protect 
user privacy than devices made by more obscure brands. 
As a result of trust-based purchasing decisions, participants were generally
confident that their IoT devices included adequate privacy protections, requiring
no additional actions for privacy preservation. 
This result extends the findings of a previous experimental study involving temporary device ownership~\cite{Worthy} by indicating that trust drives real-world purchasing and use decisions for smart home owners outside of the laboratory.

\textbf{Users are skeptical of privacy risks from devices that do not record audio or video}, such as lightbulbs and thermostats (Section~\ref{sec:findings-av}). 
They are unaware of the possibility of machine learning algorithms using non-A/V data to infer more sensitive information, including sleep patterns and home occupancy.
This evidence regarding the lack of concern about smart home data privacy risks from modern inference algorithms has not been reported in prior works.

These results provide new evidence of users' engagement with unique, IoT-specific privacy considerations related to the physical nature of these devices.
Based on these results, we suggest recommendations for designing IoT devices 
to reduce the burden of privacy controls which are mostly placed on users within the current United States consumer IoT market model. 
Device designers should seek to improve the convenience of privacy control settings on devices and associated mobile applications. Creative solutions are needed for notifying users of ongoing data collection by IoT devices without traditional screen interfaces. Additional research is also needed to develop mechanisms for centralized privacy control by smart home hubs in ways that seamlessly integrate into home life and meet users' privacy needs with minimal effort. 
The current inability of participants to discern device privacy properties also supports the creation of a 
certification program to indicate device privacy features and encourage
competition between IoT manufacturers for developing privacy-preserving devices that respect home privacy needs. 
Lowering the barrier to understanding and controlling smart home data collection will be necessary for otherwise unconcerned users 
to have privacy protections by default with minimal burden to actively configure privacy features as desired.

This paper makes the following contributions:
\begin{itemize}
    \item Supplies new evidence that users' evaluations of privacy risks from IoT devices and non-domestic entities who create, manage, track, or regulate IoT devices and data are based on stereotypical views of these entities and do not account for modern inference techniques applied to non-audio/visual data.
    \item Provides qualitative descriptions of user thought processes and rationales that confirm, explain, and extend previous survey findings about IoT privacy in home contexts. 
    \item Highlights the importance of convenience and trust as driving factors in users' decisions about incorporating IoT devices into their home and corresponding privacy concerns. 
    \item Provides new data about how user opinions of targeted advertising based on data from physical in-home devices with always-on sensors reflect a balance between perceived privacy invasion versus advertisement relevance. 
    \item Synthesizes interview responses and previous proposals into recommendations for IoT device manufacturers, researchers, regulators, and industry standards bodies that fit with user opinions and the current U.S. consumer IoT market model, which places (perhaps excessive) burden on users to manage privacy settings. 
\end{itemize}

The remainder of the paper is structured as follows:
Section~\ref{Related Work} surveys related work. Section~\ref{Method} details our
interview methods and participants.
Section~\ref{sec:limitations} describes some limitations of our participant pool and corresponding motivations for future studies.
Section~\ref{Findings} presents the major themes emerging from our
interviews. Section~\ref{discussion} discusses recommendations based on our
findings for IoT device designers, researchers, industry standards, and consumer incentives.

\section{Related Work}\label{Related Work}
The privacy implications of IoT devices are of significant interest to researchers. Smart homes demand particular attention due to societal and legal expectations of privacy in the home. 
Our study contributes to this area by identifying privacy concerns,  preferences, and behaviors of users who have independently incorporated smart home devices into their daily routines. Our results provide new evidence that users' decisions about purchasing and using smart home devices involve unique, IoT-specific privacy considerations fundamental to the physical nature of these devices. In our work, we focus on users' privacy expectations from entities external to the home who create, manage, track, or regulate IoT devices and/or their data. We recognize that there are also privacy expectations that users may have regarding how IoT devices are used and shared by household members and that these preferences may be affected by everyday routines. However, this is not a focus of our study (Section~\ref{sec:limitations-inhome}). 

Related work on smart home privacy has outlined technical and design challenges \cite{Arabo,Bugeja} and revealed potential attacks to recognize users and behaviors from data collected from smart environments \cite{Molina-Markham, Conti, Obermaier, Apthorpe, Srinivasan}. Correspondingly, many papers have suggested design and analysis frameworks \cite{Perera,Jacobsson, Nurse} and proposed systems for data management and visualization in smart environments \cite{Carminati, Wu, Mayer, Ouaddah}. The range of proposed solutions for IoT privacy include design, network, and sociotechnical efforts; however, as Jacobsson et al. \cite{Jacobsson} note, it is crucial to understand users in order to create usable privacy mechanisms. This motivates sociological research leading to greater understanding of user interaction with IoT devices.

Our work contributes to the body of research examining user opinions and interactions with the IoT and smart home environments with respect to non-household entities that create, track, and regulate IoT devices and/or data.
Due to the rarity of people living with Internet-connected home appliances before the recent surge in IoT interest, most experimental user privacy studies on IoT technologies have been conducted in temporary or laboratory settings. 
Worthy et al. \cite{Worthy} ran a week-long study of five users living with a custom IoT device. This study identifies trust as a critical factor and discuss designing IoT devices to encourage user trust. While not focused on smart homes, recent user studies investigating privacy concerns with smartwatches \cite{Udoh} and Internet-connected toys \cite{McReynolds} have also illuminated user perceptions and informed better designs for IoT device privacy. Additionally, studies of senior citizens living with monitors and sensors \cite{Townsend,Demiris}, although not Internet-connected, provide insights into user trade-offs between privacy and autonomy. Our work contributes to these studies by offering a firsthand opportunity to learn from users' long-term experiences living with a variety of IoT devices in their homes.

Survey studies have also investigated user opinions about IoT privacy. 
In 2017, Consumers International published the results of international surveys investigating consumer opinions about the growing prevalence of Internet-connected devices across industries~\cite{Consumers}. Over 60\% of the worldwide respondents reported safety concerns about connected objects.
Surveys by Choe et al. and McCreary et al. \cite{Choe,McCreary} have indicated that American users are especially concerned about Internet-connected devices recording and sharing private in-home activities. Martin and Nissenbaum \cite{Martin} surveyed 569 individuals and found that the intended use of collected data was more relevant to users' privacy opinions than the sensitivity of the data itself. This supports our finding that users weigh privacy risks against perceived benefits. Other large scale studies of IoT privacy preferences exist.
Lee and Kobsa \cite{Lee} surveyed 200 individuals to categorize IoT information collection context parameters based on respondent reactions. Similarly, 
Emami-Naeini et al.~\cite{Emami-Naeini} surveyed 1,007 participants to measure their privacy expectations in 380 IoT device use cases and scenarios.
Finally, Apthorpe et al.~\cite{Apthorpe-CI} surveyed 1,731 individuals' perceived acceptability of 3,840 information flows involving a range of IoT devices, information types, data recipients, and collection conditions. 
Each of these survey studies have demonstrated that ``privacy preferences are diverse and context-dependent''~\cite{Emami-Naeini}, supporting the importance of interview studies to uncover nuanced user opinions and elucidate the reasons why survey participants respond in particular ways to IoT privacy issues. 

Concurrently to our work, Zeng et al.~\cite{Zeng} interviewed fifteen individuals living in smart homes. They found lapses in users' understanding of IoT threat models related to the skepticism about non-A/V device privacy risks expressed by our interview participants. They also recorded potential social issues arising from multi-user smart homes, which were not expressed by our participants, motivating continued research to understand the breadth of user experiences, especially for marginalized or other at-risk individuals. Ultimately, their recommendations corroborate ours, emphasizing device design improvements to address user privacy needs.
However, our work focuses more on users' underlying value judgments and rationales for decisions about home IoT device ownership, use, and privacy management than on mental threat models of specific privacy risks.
	
Though not exclusively focused on privacy and IoT technology, a related set of research has studied the use of technologies inside the home and how their usage depends on the context of daily routines and relationships between household members. Several studies \cite{Brush,Woodruff,Woo} have investigated home automation, interviewing households to provide directions for future research and to improve the user experience with home automation. Similarly, Chetty et al. \cite{Chetty} studied networked homes to understand the relationships between households, their inhabitants, and networks to explore the potential of wireless networking in the home. In other work, Mazurek et al. \cite{Mazurek} studied access control for home data sharing, providing guidelines for usable access-control systems. More recently, Yang and Newman \cite{Yang} investigated user experiences living with a Nest learning thermostat to inform the development of intelligent home technologies. These works highlight the importance of understanding user values and behaviors and have directed the design of emerging technology for the home. These user studies guided the methods and approach of our work. 

We contribute to this body of literature by identifying perceptions of privacy and security issues for a wide variety of IoT devices from users who have independently chosen to incorporate these devices into their homes. Our focus recognizes that the notion of privacy is situated and depends on the context, everyday routines, and relationships between household members. However, we focus on how users perceive privacy issues arising from those external entities who actually create, track, regulate, or manage IoT devices and/or their data.

\section{Interview Method}\label{Method}
We conducted interviews with eleven owners of smart home devices to understand their privacy
values and expectations. The study was approved by our institution's Institutional Review Board (IRB). We recruited participants and conducted interviews from November 2016 to February 2017.

\subsection{Recruitment}
We recruited households by posting flyers in the local area, emailing listservs, and asking participants through word of mouth. In our advertisements, we did not use the words ``privacy'' or ``security'' to avoid bias in recruiting. The recruitment message requested participants with at least one of the following IoT devices in their homes: smart thermostats, lights, switches, alarms, cameras, locks, or motion sensors, but did not place any other restrictions on participation. We recruited for users with these devices because they collect a wide range of household data, allowing us to explore user attitudes towards different types of data and corresponding privacy concerns.

After receiving twelve responses to our recruiting efforts, we asked respondents to provide an inventory of their smart home devices. We selected the eight households with more than two devices in their homes. 
These eight households were selected to maximize the number of device types owned by participants within the limited time available for the study.
We then interviewed as many participants from each selected household as were willing, leading to a total of eleven participants, ranging from one to two per household. 
After interviewing these eleven participants, we reached data saturation and did not recruit further households. Our final sample may not cover a fully representative sample of IoT device users (Section~\ref{sec:limitations}). However, our interviews serve as informative case studies of early adopters of IoT home technology.

\subsection{Participants}
Household participants and compositions are detailed in Table \ref{tab:Table 1}. There were six female and five male participants  ranging from 23--45 years old. The majority of the households were from the Seattle metro area. Other households were from New Jersey, Colorado, and Texas. Participants came from a variety of living arrangements, including families, couples, and roommates. 

Our participants were fairly affluent, technically skilled, and highly interested in new technology, fitting the profile of an ``early adopter'' \cite{Rogers}. Seven of the eight households had at least one resident with a background in computer science or the technology industry, including one participant who was an expert on IoT smart home devices. We had difficulty finding households who did not fit the early adopter profile due to the relatively recent introduction of IoT smart home devices to the mass market at the time of the interviews. While we would have liked to recruit a more diverse sample, we were unable to do so within the time and constraints of our study. Having reached data saturation for our goals, we did not engage in further recruiting efforts. 

As early adopters, all households set up their own smart home devices, and we asked each household to self-identify the ``technology expert,'' or the individual who was primarily responsible for the IoT devices. During our interviews, we spoke to the self-identified ``technology expert'' of each household in order to understand their smart home setups. We also spoke to any other willing household members. 
Interviewing households together (as opposed to holding one-on-one interviews with each member separately) may have led to some hesitation from participants in terms of expressing their opinions fully. For instance, some opinions, specifically related to privacy needs relevant to relationships among household members, may not have come up in this setting (Section~\ref{sec:limitations-inhome}). Since this was not the focus of our work, we did not ask probing questions to overcome this limitation.

\begin{table*}[t]
    \begin{tabular}{ | l | p{3cm} | p{9cm} |}
    \hline
    Home & Occupants & Devices  \\ \hline
    H1 & Female roommate* (P1), female roommate* (P2) & Amazon Echo; Arlo Wireless Camera; Logitech Harmony Hub; Nest Learning Thermostat; Philips Hue Light Bulbs; Samsung SmartThings Hub, Motion Sensor, Multipurpose Sensor; TP-Link Smart Plugs
 \\ \hline
    H2 & Husband* (P3), wife (P4), daughter, son  & Amazon Echo; Honeywell Wifi Thermostat; wifi baby camera\\ \hline
    H3 & Husband* (P5), wife & Amazon Echo; Insteon Dimmer Switch and Light Switches; Nest Learning Thermostat \\ \hline
    H4 & Husband*, wife* (P6) & Amazon Echo; Arlo Security Cameras; Aeotec Siren; First Alert Z-Wave Smoke and Carbon Monoxide Alarm; Fortrezz Siren; LIFX Color and White LED Bulbs; Samsung SmartThings Hub, Open/Closed Sensor, Motion Sensor, Presence Sensor \\ \hline
    H5 & Husband, wife* (P7), daughter & Amazon Echo and Dot; August Doorbell; Belkin Wemo Switches; Cree Light; Curb; Drop Kitchen Scale; Ecobee SI; GE Light Link lights; Google Home; Kwikset Z-wave lock; Lutron dimmers; MyQ garage door opener; Nest Learning Thermostat; Netatmo; Osram Lightify; Philips Hue lights; Samsung SmartThings Hub; Sonos; UnderArmour Wi-Fi scale; Wink hub; Z-wave open-close switches\\ \hline
    H6 & Husband* (P8), wife (P9) & Amazon Echo; Dlink video cameras; LG switches, outlets, motion sensors, door sensors; Nest Learning Thermostat and Smoke Detectors; Samsung SmartThings Hub \\ \hline
    H7 & Husband, wife* (P10) & Abode Home Security System; Amazon Echo; Belkin Wemo Switches; Canary Security Camera; Gogogate garage door opener; Philip Hue bulbs; Nest Learning Thermostat and Protect; Ring doorbell; Slock smart lock; TP-Link Smart Plugs \\ \hline
    H8 & Husband* (P11), wife & Homebrite smart LED lightbulbs; x-10 smarthome switch, plug-in lamp controllers, plug-in appliance controller, motion sensor, remote controls, programmable timer\\ \hline
    \end{tabular}
    \caption{\label{tab:Table 1}Households and interview participants (* denotes self-identified technology expert of the household)}
    \end{table*}

\subsection{Interviews}
Prior to our interviews, all households filled out a pre-survey asking for demographic information, including home location, occupants, professions, and an inventory of their smart home devices (Table \ref{tab:Table 1}). Interviews were then conducted by the researcher via Skype video call.  The shortest interview lasted 30 minutes and the longest lasted 60 minutes. All interviews were recorded.  Participants were compensated with a \$50 Amazon gift card.

The interviews were structured as follows:
Participants first gave a tour of their smart homes, showing the setup of their IoT devices. We then interviewed participants about their experiences with the IoT devices in their homes. These conversations included what led them to install the devices, their favorite and least favorite aspects of the devices, 
and how they incorporate the devices into their daily routines. 
We then discussed issues related to privacy, including their awareness of what data their devices collect, their concerns about where the data goes and who has access to it, their thoughts on privacy trade-offs, and what actions they take to protect their privacy.  We asked the following specific questions in every interview:
\begin{itemize}
    \itemsep=-1pt
\item How do your devices work?
\item What data do your devices collect?
\item How would you feel if someone else knew about this data?
\item Do you have any privacy concerns about these devices?
\item Would you be willing to take any steps to protect your privacy?
\end{itemize}

While the research group prepared these specific questions in an interview guide, all interviews were semi-structured. The interviewer followed up on topics of interest that arose naturally in discussions with each household, which varied greatly depending on the composition of the household, the person interviewed, and the devices they owned. 

\subsection{Analysis}
To analyze the interviews, we first transcribed the audio recordings and analyzed participant answers to our structured questions. We then performed open coding, a process of categorizing and describing observed behaviors with ``codes'' of words or short phrases \cite{Seidman}. Codes were initially drawn from interview questions and then the themes that emerged from iterative analysis of the interviews. Parent code examples based on our interview guide include ``Demographics,'' ``Awareness,'' ``Solutions,'' and ``Concerns.'' Child code examples include ``Convenience,'' ``Current functionality,'' and ``Manufacturer\_and\_data.'' After multiple rounds of discussion and analysis with the research group, we observed a set of themes that emerged from the interview findings.

\section{Limitations}
\label{sec:limitations}
Our survey results must be considered in context of the limited number of households interviewed. While our participants' responses identify IoT privacy concerns likely shared by many others, they should not be considered representative of the entire population of IoT device users. The following limitations must specifically be acknowledged and addressed in follow-up studies.

\subsection{Our Study Focuses on External Actors, Not In-Home Privacy Threats}
\label{sec:limitations-inhome}
Since this was not a primary focus of our study, none of our interview participants raised concerns about connected devices enabling privacy violations or other malicious behaviors between individuals within a household. We therefore focused our  analysis and recommendations on external privacy threats (e.g., governments, hackers, and industry), which were of primary concern to our participants.

However, it is essential to note that this lack of concern about in-home threats is not representative of user experiences with connected devices.
For example, in June 2018, the {\em New York Times} published an article detailing
the increased prevalence of connected home devices in domestic abuse cases~\cite{NYT}.
The article discusses how consumer IoT devices can produce a power imbalance in
home environments and enable abusers to surveil and exercise control over others
in the home. This serves as
a reminder that home environments are often unsafe spaces, especially for traditionally marginalized or otherwise at-risk individuals~\cite{Chambers, Straus}. Such individuals are less likely to volunteer to participate in studies about connected devices, leaving researchers and manufacturers unaware of the domestic risks that these new technologies pose. Continued targeted research into the impact of IoT technology on in-home privacy threats is imperative to avoid empowering bad actors and placing vulnerable individuals at further risk.

\subsection{Users in the United States May Have Unique Attitudes About Privacy}
Surveys have indicated that concern about Internet-connected devices is a worldwide phenomenon~\cite{Consumers}. However, all of our interview participants live in the United States and are therefore influenced by American perspectives on privacy. IoT device users in different regions may have differing privacy concerns. For example, American users may be generally more accepting of data collection by industry versus the state, in contrast to consumers in Europe. As such, our recommendations should be interpreted in the United States context. We recommend that future studies of IoT device users in other world regions be conducted to inform further recommendations for those areas.

\subsection{Concerns About Privacy, Security, and Safety May Overlap}
Although we structured our interviews around questions about privacy, participants' responses were necessarily affected by their impressions of device security (and its influence on their safety) and the norms of the social order. These influences are especially intertwined for physical and embedded devices, and additional research would be needed to unpack the relative contributions of overlapping values to IoT users' behaviors and opinions.

\section{Results}\label{Findings}
The main themes that emerged from our interviews are as follows: First, owners of smart homes value convenience and connectedness, which dictate their privacy expectations and behaviors. Second, their opinions about who should have access to their smart home data depends on the perceived benefit. Third, trust in device manufacturers and brand reputation dictates purchasing behavior and privacy assumptions. Finally, users are skeptical of privacy risks from non-audio/visual devices. 

\subsection{Convenience and Connectedness are Priorities}
\label{sec:findings-conv}
Participants highly valued how smart home devices made their lives easier, more connected, and more convenient. When asked why they chose to purchase IoT devices for their homes, participants overwhelmingly cited convenience as a major factor, and most added that they enjoyed staying connected to their homes, families, and pets:
\begin{quote}It's cool and also convenient. I can control all my lights with my voice, which is pretty nice. (P5)\end{quote}
\begin{quote}Everything you could do in person from the actual thermostat panel you can do from your phone, so I've grown to love that pretty quickly\dots it's more convenient to be under your warm covers and not have to run downstairs and check the temperature.~(P4)\end{quote}
\begin{quote}I have a huge sense of closeness to my family by just checking in on them\dots checking in on the devices. Like, look, the lamp's on by my husband's chair, he must be sitting there reading, or look, you know, it's warm in the kitchen; if it's extra warm in there they might be cooking dinner. It's like being with my family when I'm not. (P7)\end{quote}
Values of convenience and connectedness outweighed participants other concerns about IoT devices, including obsolescence and security issues. 
\begin{quote}I think it's more likely that a lot of these things will become obsolete\dots 
If that's what happens then I have to buy another device. It still might be worth it for the convenience.~(P10)\end{quote}
\begin{quote}[The concern] is always kind of in the back of my mind because of all that IoT stuff that always goes on, and everyone says how easily hackable they are. But I think my peace of mind that I get from having them outweighs my worry of what could be potentially taken advantage of. (P6)\end{quote}
Participants were also willing to trade privacy for convenience gained from their IoT devices. These attitudes are similar to those observed in previous studies of senior residents accepting the loss of privacy in exchange for autonomy while living with sensor technologies \cite{Demiris,Townsend}.
They also re-affirm previous work on user privacy behaviors on the Internet \cite{Kang, Acquisti, Wang} in the new context of smart homes. For example, even the participant (P11) who was most concerned about privacy (and had taken action to protect his data by setting up a custom x-10 home automation system rather than use commercial IoT devices), admitted that convenience had begun to overrule his value of privacy.
\begin{quote}I'd be very interested in trying a Nest thermostat\dots I like to say I'm thoughtful about these decisions, but very often the case is that something is really convenient so I'll do it anyways even if I do have some reservations about privacy. So yeah, even though I like to say that I take those things seriously, sometimes I do just go with what's convenient. (P11)\end{quote}
\begin{quote}In terms of big data I think there's continuously going to be a trade-off, right? I would be willing to give up a bit of privacy to create a seamless experience because it makes life easier. (P8) \end{quote}
Generally, convenience and connectedness offered by smart homes were more valuable to participants than knowing and controlling where smart home data goes. 

\subsection{Opinions about Data Access Depend on Perceived Benefit from External Entities}
\label{sec:findings-benefit}
Participants had very different attitudes toward smart home data collection depending on the identity of the entity accessing the data. We specifically asked for opinions about data collection by device manufacturers, advertisers, ISPs, and the government. In all cases, the acceptability of data collection was contingent on perceived benefit to the end user. 

It is important to note that although we asked about the above entities as separate categories, there are no clear distinctions between manufacturers, advertisers, ISPs, and the government. Manufacturers of IoT devices can be first-party advertisers, such as Google placing web advertisements for Nest products, or can sell their information to third-party advertisers. ISPs can also be manufacturers, such as Comcast's line of Xfinity Home devices. Additionally, each entity may give information to the government voluntarily or though subpoena. These blurred distinctions may become problematic for user privacy, as 
participants resort to stereotypical views of external entities involved in collecting, tracking, storing, or regulating IoT data in order to appraise their privacy threats. For example, when asked about government data collection, most participants imagine a ``typical'' government, rather than a specific instance. Unfortunately, these stereotypes are of dubious value for understanding the actual privacy risks posed by the entities, and opinions formed from them are subject to change when confronted with concrete details (e.g. when thinking about the \textit{Seattle} government in particular). 

Furthermore, user perceptions of stereotyped entities external to the home are not specific to the smart home IoT context. This means that while IoT is changing the pervasiveness and granularity of in-home data collection, users are still relying on outdated, pre-IoT models of entities in the IoT ecosystem to make purchasing, privacy, and security-related decisions. 
Future studies at regular intervals will be needed to determine whether users' mental models of the following entities and associated norms evolve to incorporate IoT-specific practices over time.

\subsubsection{Manufacturers}
Participants were least concerned about manufacturers having access to the data collected by their IoT devices. Ten out of eleven participants stated that they did not mind the manufacturers of their devices collecting and analyzing data, acknowledging that it is necessary to improve the product and user experiences.
\begin{quote}If a company wants to learn more about how people are using their devices and to make improvements over the long run\dots I could see that being beneficial. (P4) \end{quote}  
Multiple participants (P8, P9, P10, P11) also expressed hopes for manufacturers to collect anonymized data in the aggregate, similar to concerns expressed in Worthy et al. \cite{Worthy}, in which users called for IoT devices to collect only de-identified, aggregated data. The sole participant (P11) who felt uncomfortable about device manufacturers collecting his data qualified that good data collection practices would rid his concern. \begin{quote}If the data were collected anonymously, then I don't think I would have a concern about it. (P11)\end{quote}
\begin{quote}I would be okay if my data was anonymized for research because I feel like they could get better by learning from how we use it. I'd opt-in if I wasn't already opt-ed in. I'm not sure if they're already doing it or not. (P8)\end{quote}
At face value, users had little concern about manufacturers accessing their smart home data to improve their products, believing that this kind of access would ultimately benefit the end user along with the assumption that the manufacturer would responsibly collect the data.

\subsubsection{Advertisers} 
Participants had differing opinions about advertisers having access to their smart home data. 
On one hand, six out of eleven participants stated that they did not mind advertisers having access to their data, for they enjoyed receiving targeted ads or better advertising experiences. As with manufacturers, they felt that they could potentially benefit from advertisers having their smart home data, so they were not concerned. \begin{quote}It feels as a consumer, if there is a better, more seamless experience to advertise to me, like I should benefit from that in some way. Or there's that relationship where if I can get something out of that, then that seems like a better deal. (P8) \end{quote}
On the other hand, the five other participants felt more hesitant about advertisers having access to their smart home data, for they did not think it could benefit them.
\begin{quote}They're no longer really doing a service for me. They're just making money off my data, and I don't like that. (P9)\end{quote}
However, participants also qualified their opinions when they expressed desires for more transparency and control: \begin{quote}Depends on the data they were going to sell. I would want to know which aspects of my relationships they were monetizing. (P7)\end{quote}
\begin{quote}I am actually more OK with selling them my data myself. I would really want to see exactly what was in the information before selling that though. (P9)\end{quote}
While we observed a split in participants' attitudes towards advertisers, the difference came down to whether they felt like they could benefit from the transaction. Participants who believed they would benefit or could control what data was collected felt more comfortable sharing their smart home data with advertisers. 
There is an extensive literature on opinions and effectiveness of targeted advertisements, but none of these studies focus on IoT~\cite{Knoll, Schumann, Martin-Kelly, Tucker}. Our study contributes data on user perspectives of advertisers with respect to IoT devices, which has not been covered in prior work. Our result also supports previous survey findings \cite{Jung, Zhu}, in which negative perceptions of privacy invasion from targeted advertisements were offset by positive effects of perceived advertisement relevance.

\subsubsection{Internet Service Providers (ISPs)} All participants stated that they would be concerned if their ISP could collect data from their smart homes. Overall, participants viewed ISPs negatively. \begin{quote}I mean I don't like them\dots nobody does. (P1)\end{quote}
Participants (P5, P11) expressed concern about their inability to prevent their ISPs from seeing their smart home data, for they believed it inevitable that ISPs could see all data from Internet-connected devices. If given the choice, however, they would rather that ISPs not see their network traffic.

All participants agreed that ISPs should not need or have access to the data from their smart home devices. They viewed the practice as invasive and unnecessary, without an obvious benefit to users. Participants (P8, P10) brought up the idea of net neutrality and their wariness of ISPs. \begin{quote}It kind of scares me for them to know it all, because I feel like with the whole conversation around net neutrality and everything I feel like they're probably trying to do terrible things. So I worry about that. (P10) \end{quote}
In this case, participants not only believed that there was no benefit in ISPs having access to their smart home data, but were also concerned that it could be detrimental.
These results corroborate previous survey findings~\cite{Apthorpe-CI} with a different U.S. sample population showing that individuals generally view data collection by ISPs as unacceptable. Our results also provide qualitative evidence explaining user distrust of ISPs observed in quantitative survey data.

\subsubsection{Government} Participants were most concerned with the government having access to their smart home data. All participants were extremely wary of any attempt by the government to infringe on their privacy, as captured by the following statement by P9: \begin{quote}I don't like the government having that type of information. I think there becomes a real slippery slope with information that goes to the government about what you do with your personal time, and I certainly know the argument ``well you have nothing to hide then why does that matter,'' but I think things get really dicey really easily when it comes to people's personal privacy and their civil liberties. (P9) \end{quote} 
These opinions are in keeping with increased awareness and unease towards government data collection after the Edward Snowden leaks~\cite{Dencik}.
Most participants agreed that their data should be protected from government access; participants P7 and P10 also brought up concerns about subpoenaing data from smart homes: \begin{quote}I think there's a precedent. You know there was a case where a judge tried to subpoena Amazon because they thought it would help in a murder case or something like that and Amazon said no\dots I think those companies have a big incentive that they really need to make sure that the privacy is protected otherwise people won't use it [the device]. (P10)\end{quote}
Interestingly, P10 also noted that the local government could possibly improve local services by having access to smart home data: 
\begin{quote}I live in Seattle and I think that they're pretty innovative on what they're doing on a city level to understand lots of different things like what they're doing in your community and what kind of things help\dots if they're using it to try to understand electricity usage so that they could make it cheaper for everyone, I feel like that's good for society and that's fine. (P10)\end{quote} 
Again, if there seems to be a benefit for the end user, then sharing data seems to be permissible. However, most participants believed that the government would only use the data for persecution, so they were concerned about data privacy as a civil liberty. 
This conflict between perceived violations of civil liberties and perceived benefits of government data collection aligns with previous survey results measuring Americans' opinions about government surveillance of low-income populations in non-IoT contexts~\cite{Turow}.

Unfortunately, users in our study experienced blurred distinctions between manufacturers, advertisers, ISPs, and the government when it came to IoT devices and data collection, tracking, storing, and transfer. This  blurred distinction between IoT device manufacturers and other external entities has implications for user privacy, as we describe below.

\subsection{Trust in Manufacturer Privacy Protections} 
\label{sec:findings-trust}
Trust in companies was central to participants' privacy behaviors, similar to previous findings by Worthy et al. \cite{Worthy} about users living temporarily with an IoT device.
Trust in brand reputation especially influenced participants' selection of  IoT devices for their homes. All participants claimed that they did significant research before purchasing any of their smart home devices and ultimately selected devices  based on online reviews and brand reputation. 
\begin{quote}I try to make sure it's a reputable brand\dots I don't try to just get the cheapest thing because those are less secure...I want something that's more established so I kind of assume that they have more security. (P10)\end{quote} 
\begin{quote} I really wanted a [Google Home]\dots I feel like more work went into it\dots and it came from a renowned company. (P1) \end{quote}
Participants tended to trust big technology companies 
to have the technical means to protect their data, although they could not confirm if these companies actually performed encryption or anonymization. 
\begin{quote}I trust for example Amazon and Google to have good security practices so like my voice recordings and things like that\dots I don't mind that those speakers are constantly aware. (P7) \end{quote} 
Many participants also placed trust in home appliance and lighting brands, despite the fact those those companies have limited experience making Internet-connected products and the participants took no action to verify the companies' privacy protection practices. 
\begin{quote}Philips and Belkin are a little more household names so a little easier to trust. (P10) \end{quote}
\begin{quote}Third parties may not be as secure or may not have an infrastructure that's as secure as Insteon. (P5) \end{quote}
However, regardless of the size or technological capability of the company, participants overwhelmingly stated that they trusted the brands of the devices that they had chosen for their homes. 
Participants rationalized their reluctance to take extra action to protect their privacy by referring to their trust in IoT device manufacturers to not do anything malicious with their data.
\begin{quote}I assume that these things are good enough. I trust them\dots I kind of assume [protection] is already there. Like I assume the hubs are doing some magic. (P2)\end{quote} 
\begin{quote} I think it's anonymized enough where I'm not worried that somebody knows exactly what I'm doing. It's more like as a cohort or you know as a statistic so it doesn't really make me feel like I'm losing privacy or something. (P10)\end{quote} 
However, it is unclear whether the users would take serious control over their privacy if this trust in manufacturers was breached.  Given their desire for convenience and the work required to manually configure device privacy settings, users might instead decide to trust other institutions, such as regulatory agencies, to protect their privacy.

\subsection{Skepticism of Non-A/V Device Privacy Risks}
\label{sec:findings-av}
While participants were generally concerned about the privacy implications of smart home devices that record audio or video, several expressed skepticism that non-A/V devices posed privacy risks. 
\begin{quote}I'd be happier if someone hacked a plug than the camera pointed in the living room.~(P1)\end{quote}
\begin{quote}I'm not really too concerned about lights being on or off, or if someone does somehow hack my account, or if someone is able to turn my lights on or off.~(P5) \end{quote}
In particular, participants were unaware of the potential for machine learning algorithms to infer sensitive information from otherwise innocuous data, such as in-home temperatures and when the front door opens. 
\begin{quote}You know the door opened 15 times\dots and you can see how cold it is in the house. I guess that doesn't really bother me if they keep that.~(P6) \end{quote}
\begin{quote}There's nothing that a thermostat knows except what temperature our house is at at different times of the day and year, and I would not feel very private about that information.~(P3)\end{quote}
However, researchers have demonstrated that metadata from non-A/V smart home devices can provide enough information to infer user activities~\cite{Srinivasan, Apthorpe}, such as home occupancy, work routines, and sleeping patterns. 
By dividing their conceptions of smart home data into binary ``sensitive'' and ``nonsensitive'' categories without fully understanding the data analysis capabilities of entities that collect the data, participants are inadvertently jeopardizing their privacy while believing they have no cause for concern. 
This is a new contribution regarding user opinions of IoT data privacy risks from modern inference algorithms.

\section{Recommendations}\label{discussion} 

Our interview results have various implications for the designers of future smart home IoT devices, as well as for researchers, industry standards bodies,
consumer advocacy groups, and regulators.
In this section, we discuss some recommendations based on our findings.
Some of these recommendations echo results from previous work (e.g., privacy
controls must be clear and convenient). Previous studies on privacy in IoT and other contexts have proposed similar recommendations, especially for device designers~\cite{Zeng, BITAG}. However, current devices still do not follow these recommendations, so they deserve to be re-iterated. In addition, we make several novel recommendations to improve the state of user privacy with IoT devices.

It is generally accepted that users dislike spending the time and effort to manage privacy settings or other privacy controls or have difficulty configuring them to match their intentions~\cite{Madejski}. Thus, even with improved IoT privacy controls backed by improved design and stricter regulation, many users may be unwilling to take meaningful steps to manage their privacy. This motivates a larger debate about how society should distribute the burden of privacy control between users, regulators, and industry, which is out of scope for this work. In the interim, we focus on recommendations that would help ease the burden on IoT device users attempting to control their privacy and make it more likely that they will do so effectively with little effort. This presumes that the user is a rational actor who can play a part in taking control of his or her privacy using tools provided by device manufacturers within the constraints of government regulation. We do not however presume it is solely the responsibility of the user to do so.

\subsection{Smart Home Device Designers}
Participants' desire for convenience and general trust in IoT device manufacturers
limits their willingness to take action to verify or enforce smart home 
data privacy. 
This reality means that privacy notifications and settings must be 
exceptionally clear and convenient if they are to be used. 
It is well-known that user-friendliness increases engagement with privacy settings and warnings
in web and mobile application contexts~\cite{Egelman, Egelman-Trust}; 
however, most smart home IoT devices do not have screens. This makes it considerably more 
difficult for designers to incorporate privacy notifications and settings options that are clear, simple, and accessible
without overwhelming the user interface of the device~\cite{Regal}.

For some forms of data collection, clear non-screen notifications already exist. The most recognizable 
is likely the light that turns on next to a camera (e.g., on a laptop) when it is recording. IoT 
personal assistants (Google Home and Amazon Echo) already display particular 
light patterns to indicate that the device is recording audio. 
Since interview participants expressed greater privacy concern about devices
that record voice and video, we recommend that such visual indicators
be used extensively to indicate these activities, especially in devices 
traditionally without recording capabilities (e.g. doorbells, lightbulbs, etc.). Previous researchers have conducted extensive work to determine the effectiveness of webcam indicators as well as security connection indicators~\cite{Portnoff, Machuletz, Shi}. We propose that future work focus on what indicators would work for different kinds of IoT devices so as not to overwhelm users.

Another option is for privacy notifications and settings to be incorporated into 
mobile applications associated with smart home devices. 
Interview participants primarily interacted with their IoT devices through mobile
applications, especially the applications of hub-style devices that connect and control 
multiple other devices in home.
We recommend that the mobile applications for IoT devices have privacy
settings similar to those commonly available on web applications, 
allowing users to see and control which data their devices can
collect and send to the cloud. 
The applications should allow users to view and delete data points, such
as mistaken voice commands or any kind of sensitive material. 
A few participants mentioned that they appreciated the ability to use the
Amazon Echo application on their phones in order to see recorded commands and delete
those of their choosing. We recommend similar practices for all IoT device mobile applications.

Some researchers
\cite{Sivaraman,Simpson} have proposed building dedicated hubs for privacy and security
control in smart home networks. However, we believe that it is
unlikely that users would purchase a dedicated device to protect their smart home data, 
especially if it requires additional effort to set up and maintain without readily tangible 
benefits.
Ultimately, there is an underlying tension between convenience---a primary
reason why our participants purchased IoT devices---and privacy control, which,
if left to the user, necessitates at least some additional effort. All practical privacy solutions will have to operate within this constraint. We therefore look to researchers, regulators, and industry standards for alternative approaches to reduce the burden of privacy control on users.

\subsection{Researchers}
Even with the previous design recommendations, mobile application privacy settings and simple visual indicators will not be effective for all smart home devices. 
In homes with more than
a few devices, it is overwhelming to have separate mobile applications with
independent privacy settings for each device.  
Mobile application privacy settings will also be ineffective for non-A/V devices if users do not believe that the privacy risks justify time and effort to modify default settings.  
Additionally, many devices warrant
on-device privacy notifications more nuanced than a single light indicator. 
Researchers have a significant part to play investigating the current 
challenges users face in managing privacy and security for their growing number of home devices. 

Studies are needed to help untangle the underlying sensitivities around different domestic practices that are detected and recorded by connected devices. Tolmie et al.~\cite{Tolmie} note that while some types of data are innocuous under most circumstances, specific information about a household (known by external actors or household members themselves) may allow compromising inferences from the data. As these synergistic combinations of data and information are difficult to anticipate, researchers should combine user studies with data collection measurements
to develop methods of detecting data streams
that may violate user privacy \textit{in situ}.
These methods could then be used to design interfaces that will make users more aware of unexpected data collection behaviors and ways to align their privacy and security needs with controls that are not overwhelming. Possible solutions include creating systems that make it easy to centralize management of privacy and security across home devices, helping users by providing the strongest privacy and security settings by default, or providing recommendations as to what settings best match users' unique domestic environments. Crowdsourcing or otherwise automating suggested defaults for different situations or users could arguably decrease the need for users to customize their privacy and security settings~\cite{Lin, Ausloos}.

Hub devices and personal assistants, such as the Amazon Echo, Google Home, and Samsung SmartThings hub, already allow users to control many smart home devices from one centralized interface. 
These devices could also serve as centralized points of privacy control for the entire smart home network. 
However, this would require agreement on privacy-related APIs supported by devices in the ecosystem. 
The format of such APIs is an open research question, as privacy is more nuanced than the functional APIs required for controlling specific device features.
One could imagine a permissions model similar to the Android operating system, where smart home hubs request and monitor permissions for other IoT devices on the network. The available permissions could include which sensors are allowed to be active and which entities are allowed to receive specific data (e.g., ``only the manufacturer can receive data derived from temperature recordings'').
Additionally, devices would need to allow meaningful settings options without breaking functionality. If the only options are ``no network access'' (and therefore no Internet-related features) and ``unlimited network access,'' users prioritizing connectedness and convenience will not choose the restrictive option.
Additional technical and HCI research is needed to develop simple and meaningful privacy settings and corresponding enforcement mechanisms for centralized smart home privacy control. 

\subsection{Regulation, Industry Standards, and Consumer Incentives}

As discussed in Section~\ref{Findings}, distinctions between categories
of smart home service providers are becoming increasingly blurred. From
application to service to content providers, the
IoT market is proliferating, but not all companies are regulated by the same
agencies. Different device manufacturers may therefore be held to different technical or
privacy standards. For example, Comcast is currently regulated as a common
carrier by the Federal Communications Commission (FCC); the Federal Trade
Commission (FTC) cannot pursue or enforce regulation with respect to any
Comcast products, including Xfinity Home, a suite of home automation and security products. Yet, companies such as Google and Samsung, who
manufacture essentially the same suite of products (e.g., Google Home, Samsung
SmartThings) are subject to FTC regulatory action. These
differences are especially troubling given that users have different
perceptions about these companies, leading to vastly different privacy opinions
for essentially the same types of devices.

The European Union's General Data Privacy Regulation (GDPR) and the the California Consumer Privacy Act of 2018 (AB 375)~\cite{NYT-Cal} are examples of how legislative action can institute more uniform privacy protections across the industry. However, it remains to be seen whether these laws will have an appreciable impact on the data collection behavior of consumer IoT devices or address consumers' misgivings about IoT device privacy.
Such regulations will have especially little influence on issues of interpersonal privacy regarding IoT data accessible to multiple household members~\cite{Goulden}. Follow-up studies after these laws have been in effect for a longer period will be necessary to judge whether IoT users are satisfied with their results.

Another possibility is to create a certification program with a common set of industry standards
outlining best privacy practices for
for device manufacturers. Consumers could then evaluate the
privacy practices of a device or manufacturer based on adherence to this common
set of guidelines.  Such standards could determine, for example, to what
extent various devices send anonymized, encrypted, and aggregated data to
different entities. Products that do meet privacy standards could be highlighted as
such. Williams et al. \cite{Nurse} report that privacy standards would further
encourage competition in a market already replete with insecure devices.
Perera et al. \cite{Perera} have suggested a framework to assess the privacy
capabilities and gaps of IoT devices, such as whether collection of data is
minimized, anonymized, encrypted, distributed.

However, standardizing privacy practices
poses significant challenges. 
There is a risk, for example, in being overly prescriptive about specific technologies or practices, lest a
certain technology prove to be insufficient for protecting consumer privacy.
Finally, consumers must  care about the certification, requiring advertising and outreach 
to inform customers of privacy risks and the safeguards provided by certified devices.  
\section{Conclusion}\label{Conclusion}

Our interviews of smart home owners provide insights into long-term
experiences living with IoT devices. Recurring themes indicate that users prioritize convenience and connectedness,
and these values dictate their privacy opinions and behaviors.
User opinions about who should have access to their smart home data depend on perceived benefit from entities external to the home that create, track, regulate, or manage IoT devices and their data. Users also assume their privacy is protected based on
trust in  IoT device manufacturers, but are unaware of the potential for
machine learning inference to reveal sensitive information from
non-audio/visual data.
These findings provide new evidence of users' IoT-specific privacy considerations and suggest the need for improved privacy notifications
and user-friendly settings, as well as industry privacy standards that cut across
regulatory divisions.  As IoT home devices become increasingly
ubiquitous, our study's findings and recommendations
contribute to the broader understanding of users' evolving attitudes
towards privacy in smart homes.

\begin{acks}
We thank our study participants. This work was funded by NSF CNS-1539902.
\end{acks}

\bibliographystyle{ACM-Reference-Format}
\bibliography{SmartHomeUserPrefs}

\end{document}